\documentclass[conference]{IEEEtran}
\ifCLASSINFOpdf
\else
\fi
\usepackage{amsmath,amssymb,amsthm}
\usepackage{color}  
\usepackage{accents}
\usepackage{comment}
\usepackage{subcaption}
\usepackage[export]{adjustbox}
\usepackage{graphicx}
\usepackage{hyperref}

\newcommand\diag[1]{\mathrm{Diag}\{#1\}}
\newcommand\norm[1]{\left\lVert#1\right\rVert}

\newcommand{\x}{\mathbf{x}}
\newcommand{\y}{\mathbf{y}}
\newcommand{\R}{\mathbb{R}}
\newcommand{\V}{\mathcal{V}}
\newcommand{\A}{\mathbf{A}}
\renewcommand{\L}{\mathbf{L}}

\newtheorem{theorem}{Theorem}
\newtheorem{lemma}{Lemma}

\newtheorem{proposition}{Proposition}
\newtheorem{ass}{Assumption}
\usepackage[style=ieee]{biblatex}
\begin{document}
%
\title{A Decoupled Approach for Composite Sparse-plus-Smooth Penalized Optimization}

\author{\IEEEauthorblockN{Adrian Jarret}
\IEEEauthorblockA{\textit{LCAV, EPFL}}
\and
\IEEEauthorblockN{Valérie Costa}
\IEEEauthorblockA{\textit{M.Sc. Student, EPFL}}
\and
\IEEEauthorblockN{Julien Fageot}
\IEEEauthorblockA{\textit{LCAV, EPFL}}}


%


\maketitle

\begin{abstract}
We consider a linear inverse problem whose solution is expressed as a sum of two components: one smooth and the other sparse. This problem is addressed by minimizing an objective function with a least squares data-fidelity term and a different regularization term applied to each of the components. Sparsity is promoted with an $\ell_1$ norm, while the smooth component is penalized with an $\ell_2$ norm.

We characterize the solution set of this composite optimization problem by stating a Representer Theorem. Consequently, we identify that solving the optimization problem can be decoupled by first identifying the sparse solution as a solution of a modified single-variable problem and then deducing the smooth component.

We illustrate that this decoupled solving method can lead to significant computational speedups in applications, considering the problem of Dirac recovery over a smooth background with two-dimensional partial Fourier measurements.
\end{abstract}


%
\IEEEpeerreviewmaketitle

\section{Introduction}
\label{sec:intro}

We consider the composite \textit{sparse-plus-smooth} optimization problem, defined as
\begin{equation}
\begin{split}
        \underset{\x_1, \x_2 \in \R^N} {\arg\min} \frac{1}{2} & \norm{\y - \mathbf{A}(\x_1 + \x_2)}_2^2 \\ &+ \lambda_1 \norm{\L_1\x_1}_1 + \frac{\lambda_2}{2} \norm{\L_2\x_2}_2^2.
\end{split}
\label{eq:composite-pbm}
\tag{$P_{12}$}
\end{equation}

This problem is used to find a solution to the linear inverse problem $\y = \A (\x_1 + \x_2)$ with forward matrix $\A \in \R^{L \times N}$ and measurements $\y \in \R^L$, while decomposing the solution into the sum of two components with different characteristics. The problem \eqref{eq:composite-pbm} involves two \textit{generalized} regularization terms, associated to the penalty matrices $\L_1 \in \R^{M_1 \times N}$ and $\L_2 \in \R^{M_2 \times N}$, that encode the properties of each recovered component. The parameters $\lambda_1 > 0$ and $\lambda_2 > 0$ control the intensity of the regularizations.

\subsection{Motivation}

Using an optimization problem involving two components with different characteristics is a versatile way to model heterogeneous signals. This strategy improves upon the classical single-component techniques, such as the (generalized) Ridge or LASSO regressions \cite[Chapter~3.4]{hastie2001}, which only handle signals with a homogeneous structure. Hence, the composite problem \eqref{eq:composite-pbm} can be seen as a tool to enforce a fine prior model on the solution of the inverse problem $\y = \A \x$, while separately accessing the two components of this model.

Many works have focused on the reconstruction of composite signals, with early applications in image reconstruction \cite{osher2003,daubechies2005,starck2005} and high frequency denoising \cite{marziliano2006}. Recent works have converged to the formalism of regularized optimization to solve linear inverse problems, leading to the problem \eqref{eq:composite-pbm} that can be found in applications \cite{mol2004,gholami2013a,debarnot2021a} and which is usually solved with coupled proximal methods \cite{gholami2013a,briceno-arias2011,naumova2014}.


In this communication, we consider the composite reconstruction of sparse-plus-smooth images in a setup that mimics a radio interferometry imaging problem in astronomy \cite{thompson2017a}. Indeed, the celestial objects observed in radio astronomy can be modelled as sparse point sources over a dense smooth background, and the measurement operator of radio interferometry is generally assumed linear. The optimization problem~\eqref{eq:composite-pbm} then seems well-suited to recover the inherent composite structure of astronomical images.

The problem \eqref{eq:composite-pbm} can be solved using proximal methods, such as the Accelerated Proximal Gradient Descent algorithm (APGD) \cite{liang2017}, which demonstrates a theoretically optimal convergence rate. However, this approach solves a problem of size $2N$ whose variable is the concatenation of the components $\x_1 \in \R^N$ and $\x_2 \in \R^N$. Additionally, such a coupled solution to the composite problem rules out the use greedy atomic methods, although these methods are known to be numerically efficient for sparse problems \cite{jarret2022,Jaggi_2013}.

\subsection{Our contributions}

Our main result, Theorem~\ref{thm:rt}, is a Representer Theorem (RT) that characterizes the solution set of the composite problem~\eqref{eq:composite-pbm}. 

The RT identifies that the sparse and smooth component, respectively $\x_1$ and $\x_2$, are naturally disentangled. For any regularization parameters $\lambda_1, \lambda_2 > 0$, the sparse component solution $\x_1^*$ is determined as the solution of a convex problem with an $\ell_1$-based regularization term, that does not involve the smooth component $\x_2$. Even though there might exist many solutions $\x_1^*$, the smooth component solution $\x_2^*$ is itself always unique. Solving the minimization problem~\eqref{eq:composite-pbm} of size $2N$ can then be decoupled into solving two problems of size $N$.

We illustrate that this decoupled approach can significantly reduce the time required for numerical minimization of the problem, recovering the same solution as the coupled approach. 

\subsection{Representer Theorems for penalized optimization problems}
Our main result is analogous to several RT already present in the literature.
The single-component problem over $\x_2$ with the generalized \textit{Tikhonov} penalty term $\frac{\lambda_2}{2}\norm{\L_2 \x_2}^2_2$ admits a unique closed-form solution under mild assumptions \cite[Chapter~5.1]{Hansen_1998}. Regarding the generalized LASSO problem resulting from the single-component problem $\x_1$, the solution set is characterized in \cite{unser2016} for invertible matrices $\L_1$ and in \cite{flinth2019a} in the general case.

Composite reconstruction over sums of Banach spaces is treated in \cite{unser2022}. In particular, sparse-plus-smooth problems over continuous-domain signals are studied in \cite{debarre2021}. The authors demonstrate that each component satisfies a RT for the corresponding single-variable optimization problem with the associated penalty. Our result completes their RT by providing a more detailed expression of the solutions for discrete problems, which highlights the decoupling between the components.

\subsection{Notation}
The adjoint of a real-valued matrix $\A \in \R^{N \times L}$ is the transpose matrix $\A^T \in \R^{L \times N}$. The nullspace of $\A$ is denoted as $\ker{\A}$, and its orthogonal complement as $\ker{(\A)}^\perp$, so that $\ker{\A} \oplus \ker{(\A)}^\perp = \R^N$. The problems and results in Section~\ref{sec:intro} and Section~\ref{sec:rt} are stated for real-valued matrices and input vectors $\x \in \R^N$, but hold for any finite-dimensional inputs. In particular, the problem studied in Section~\ref{sec:application-case} involves input images $\x \in \R^{n\times n}$ and complex-valued measurements $\y \in \mathbb{C}^L$. With a slight abuse of notation, we use the Hermitian transpose notation $\A^H : \mathbb{C}^L \to \R^N$ to denote the adjoint operation in this situation, that results from the bijective mapping $\mathbb{C} \approx \R^2$ and the real-valued inner product $\langle \mathbf{z}_1, \mathbf{z}_2 \rangle_\mathbb{C} = \Re \left( \mathbf{z}_1^H \mathbf{z}_2 \right)$ \cite[Section~7.8]{Rimoldi_2016}. The numerical processing is nonetheless conducted with real-valued operations. 

For any vector $\x_1, \x_2 \in \R^N$, the notations $\x_1 \odot \x_2$, $\x_1^2 = \x_1 \odot \x_1$ and $\x_1 / \x_2$ are used for pointwise operations. The circular convolution between two vectors appears as $\x_1 \circledast \x_2$. The identity matrix of size $N$ is denoted as $\mathbf{I}_N$ and $\mathbf{1}_N$ is the vector of ones of size $N$. The operator $\diag{\cdot} : \R^N \to \R^{N \times N}$ creates a diagonal matrix from an input vector.


\section{Decoupling with a New Representer Theorem}
\label{sec:rt}

We first consider the following three assumptions:


\begin{ass}
    \label{ass:1}
    The forward matrix $\mathbf{A} \in \R^{L \times N}$ is surjective, \textit{i.e.}, has full row rank, so that $\A\A^T$ is invertible.
\end{ass}

\begin{ass}
    \label{ass:2}
    The nullspaces of the forward matrix and the regularization matrix $\L_2 \in \R^{M_2 \times N}$ have a trivial intersection, that is $\ker{\A} \cap \ker{\L_2} = \{\mathbf{0}\}$ .
\end{ass}

\begin{ass}
    \label{ass:3}
    The vector space $\ker{(\A)}^\perp$ is an invariant subspace of the operation $\L_2^T\L_2$, \textit{i.e.}, the following holds: $\x \in \ker{(\A)}^\perp \Rightarrow \L_2^T\L_2\x \in \ker{(\A)}^\perp$.
\end{ass}

\noindent Under Assumption~\ref{ass:1}, we define the matrix $\mathbf{\Lambda_2}\in \R^{L \times L}$ as
\begin{equation}
    \label{eq:Lambda2}
    \mathbf{\Lambda_2} = \left(\mathbf{A}\mathbf{A}^T\right)^{-1} \mathbf{A} \L_2^T \L_2 \mathbf{A}^T.
\end{equation}
We can now state our main result, the proofs of which are deferred to the end of this section.

\begin{theorem}[RT for the composite problem \eqref{eq:composite-pbm}] \label{thm:rt}
    Under Assumptions~\ref{ass:1}, \ref{ass:2} and \ref{ass:3}, the solution set $\V$ of \eqref{eq:composite-pbm} can be written as the Cartesian product 
    \begin{equation*}
        \V = \V_{1} \times \V_{2} \vspace{-12pt}
    \end{equation*}
    where :
    \begin{enumerate}
        \item The sparse variable $\x_1$ belongs to the set $\V_{1}$ defined as        
            \begin{equation}
                    \hspace*{-.3cm}\V_{1} = \underset{\x_1 \in \R^N}{\arg\min} \, \left\{
                     \begin{aligned}
                    \frac{1}{2} \left(\y - \mathbf{A}\x_1 \right)^T & \mathbf{M}_{\lambda_2} \left(\y - \mathbf{A}\x_1 \right)  \\
                    &  + \lambda_1 \norm{\L_1\x_1}_1
                    \end{aligned}
                    \right\}
            \label{eq:decoupled-sparse}
            \tag{$P_1$}
            \end{equation}
            with $\mathbf{M}_{\lambda_2} = \lambda_2 \mathbf{\Lambda_2} \left( \mathbf{AA}^T + \lambda_2\mathbf{\Lambda_2} \right)^{-1} $;
        \item All the sparse component solutions share the same measurement vector, that is there exists $\tilde{\y}\in \mathbb{C}^L$ such that any $\x_1^* \in \V_1$ satisfies $\mathbf{A}\x_1^* = \tilde{\y}$ ;
        \item The smooth component solution is unique and independent of the sparse component, so that $\V_2 = \left\{ \x^*_2 \right\}$. $\x_2^*$ is the unique solution of the minimization problem
            \begin{equation}
            \begin{split}
                    \underset{\x_2 \in \R^N}{\arg\min} \frac{1}{2} & \norm{\y - \tilde{\y} - \mathbf{A}\x_2}_2^2 + \frac{\lambda_2}{2} \norm{\L_2\x_2}_2^2,
            \end{split}
            \label{eq:decoupled-smooth}
            \tag{$P_2$}
            \end{equation}
        given by $\x^*_2 = \mathbf{A}^T \left(\mathbf{A}\mathbf{A}^T + \lambda_2 \mathbf{\Lambda_2}\right)^{-1} \left(\y - \tilde{\y}\right)$.
    \end{enumerate}
\end{theorem}

In applications, Assumption~\ref{ass:1} and \ref{ass:2} are commonly satisfied but it is generally not the case for Assumption~\ref{ass:3}. This latter is more restrictive and limits the scope of applications. Additionally, the inversion of $( \mathbf{AA}^T + \lambda_2\mathbf{\Lambda_2})^{-1}$ can be expensive or even intractable.
Two cases trivially satisfy the assumptions: signal decomposition with $\mathbf{A} = \mathbf{I}_N$ and simple smooth penalty $\mathbf{L}_2 = \mathbf{I}_N$. More insightful valid scenarios may come from $\mathbf{A}$ and/or $\mathbf{L}_2$ being (circular) convolutions as numerical processing and matrix inversions simplify.

Nevertheless, when applicable, the decoupled approach suggested by Theorem~\ref{thm:rt} generally simplifies solving the composite problem \eqref{eq:composite-pbm} compared to a direct coupled approach, improving reconstruction performance in runtime and quality.

The proof of Theorem~\ref{thm:rt} relies on the following lemma:

\begin{lemma} \label{lem:lemma}
    If Assumptions \ref{ass:1}, \ref{ass:2} and \ref{ass:3} hold, $\mathbf{\Lambda_2}$ satisfies
    \begin{equation}
        \left(\mathbf{A}^T\mathbf{A} + \lambda_2 \L_2^T \L_2 \right)^{-1} \mathbf{A}^T = \mathbf{A}^T \left(\mathbf{A}\mathbf{A}^T + \lambda_2 \mathbf{\Lambda_2}\right)^{-1}.
    \label{eq:prop-la}
    \end{equation}
\end{lemma}

\begin{proof}[Proof of Lemma~\ref{lem:lemma}]
    We first prove that the matrix $\mathbf{\Lambda_2}$ satisfies 
    \begin{equation}
        \mathbf{A}^T \left(\mathbf{A}\mathbf{A}^T + \lambda_2 \mathbf{\Lambda_2}\right) = \left(\mathbf{A}^T\mathbf{A} + \lambda_2 \L_2^T \L_2 \right)\mathbf{A}^T.
        \label{eq:la2}
    \end{equation}
    It then suffices to prove the invertibility of the inner terms to obtain \eqref{eq:prop-la}.
    
    Let $\mathbf{P} = \A^T\left(\mathbf{A}\mathbf{A}^T\right)^{-1} \mathbf{A}$ be the orthogonal projection onto $\ker{(\A)}^\perp$, we have $\A^T \mathbf{\Lambda_2} = \mathbf{P} \L_2^T \L_2 \mathbf{A}^T$. As $\mathrm{im}(\A^T) = \ker{(\A)}^\perp$, for any $\y\in\R^L$, we have $\A^T\y \in \ker{(\A)}^\perp$. Hence, Assumption~\ref{ass:3} implies that $\mathbf{P} \L_2^T \L_2 \mathbf{A}^T = \L_2^T \L_2 \mathbf{A}^T$ and \eqref{eq:la2} holds.
    
    Let $\mathbf{U} = \mathbf{A}^T\mathbf{A} + \lambda_2 \L_2^T \L_2$, appearing in the right-hand side of \eqref{eq:la2}. For any $\x \in \R^N$ we have $\x^T \mathbf{U} \x = \norm{\A\x}_2^2 + \lambda_2 \norm{\L_2\x}_2^2 \geq 0$. By Assumption~\ref{ass:2}, the nullspace of $\mathbf{U}$ is trivial and reduced to $\ker\mathbf{U} = \left\{ \mathbf{0}_N \right\}$. As an injective square matrix, $\mathbf{U}$ is hence invertible. Similarly, define $\mathbf{V} = (\A \A^T)^2 + \lambda_2 \mathbf{A} \L_2^T \L_2 \mathbf{A}^T$ that is also invertible. Thanks to Assumption~\ref{ass:1}, the left-hand matrix $\mathbf{A}\mathbf{A}^T + \lambda_2 \mathbf{\Lambda_2} = (\A\A^T)^{-1}\mathbf{V}$ is finally invertible and we state equation~\eqref{eq:prop-la}.
\end{proof}

\begin{proof}[Proof of Theorem~\ref{thm:rt}]
    We observe that the initial problem~\eqref{eq:composite-pbm} is equivalent to
    \begin{equation*}
        \begin{split}
            \underset{\x_1 \in \R^N}{\min} \biggl\{ \underset{\x_2 \in \R^N}{\min} & \frac{1}{2} \norm{\y - \mathbf{A}(\x_1 + \x_2)}_2^2 \\
            & + \frac{\lambda_2}{2} \norm{\L_2 \x_2}_2^2 \biggr\} + \lambda_1 \norm{\L_1\x_1}_1.
        \end{split}
    \end{equation*}
    Given Assumption~\ref{ass:1}, the inner problem over the component $\x_2$ can be solved for any $\x_1\in\R^N$. We first compute the gradient of its differentiable objective function and set it to zero
    \begin{equation}
        \A^T(\A(\x_1 + \x_2) - \y) + \lambda_2 \L_2^T\L_2 \x_2 = 0 .
        \label{eq:grad}
    \end{equation}
    Using Assumption~\ref{ass:1}, we obtain a solution $\x^*_2$ with a closed-form expression depending on $\x_1$ as
    \begin{equation}
        \x_2^* = - \left(\mathbf{A}^T\mathbf{A} + \lambda_2 \L_2^T \L_2 \right)^{-1} \mathbf{A}^T \left(\A\x_1 - \y\right).
        \label{eq:x2-from-x1}
    \end{equation}
    Using Lemma~\ref{lem:lemma}, we plug the value of $\x_2^*$ from \eqref{eq:x2-from-x1} into \eqref{eq:grad}, so that we can express the composite data-fidelity term with the unknown $\x_1$ only.
    \begin{equation}
        \A(\x_1 + \x_2^*) - \y = \lambda_2 \mathbf{\Lambda_2} \left( \mathbf{AA}^T + \lambda_2\mathbf{\Lambda_2} \right)^{-1} (\A\x_1 - \y)
        \label{eq:data-fid}
    \end{equation}
    Introducing the matrix $\mathbf{M}_{\lambda_2}$ from the theorem, we plug \eqref{eq:data-fid} and \eqref{eq:x2-from-x1} obtained that way into the problem~\eqref{eq:composite-pbm}, which leads to $\x_1$ being a solution of the $\ell_1$-penalized optimization problem~\eqref{eq:decoupled-sparse}.

    Even though \eqref{eq:decoupled-sparse} is not exactly a LASSO problem due to the matrix $\mathbf{M}_{\lambda_2}$, the data-fidelity term is still strictly convex. It is then possible to extend the result of the uniqueness of the fit $\A\x_1^*$ with a proof similar to, \textit{e.g.}, the case of the generalized LASSO problem \cite[Lemma~1.ii)]{ali2019a}.

    The uniquess of the fit $\A\x_1^*$ ensures the uniqueness of the component $\x_2^*$. The final expression for $\x_2^*$ provided in the theorem arises using Lemma~\ref{lem:lemma}.
\end{proof}

\section{\texorpdfstring{Special Case: Fourier Measurements and Convolution Operator $\L_2$}{Special Case: Fourier Measurements and Convolution Operator L2}}
\label{sec:application-case}

The decoupling of problem \eqref{eq:composite-pbm} allows to split a size $2N$ optimization problem into a problem of size $N$ and some simple matrix multiplications to compute the smooth component. However, this comes at the cost of computing the matrix $\mathbf{M}_{\lambda_2}$, which might be numerically demanding in the general case. However, in specific scenarios this computation is fairly simple and the decoupling approach is straightforward, given that the aforementioned assumptions hold.

The \textit{random Fourier measurements} model that we consider here falls into that category. The cogram matrix $\A\A^H$ is diagonal (it is even a homothety)\footnote{The matrix $\A$ is complex-valued here thus its adjoint operation is denoted with the superscript $\A^H$ instead of $\A^T$.}. When the operator $\L_2$ is described by a (circular) convolution, the matrix $\mathbf{\Lambda_2}$ is also diagonal so that $\mathbf{M}_{\lambda_2}$ can be explicitly computed with little effort. Let us present how the operations simplify in this context.

Let the indices $\mathcal{I} = \{i_1, \dots, i_L\}$ be a subset of size $L$ of $\{0, \dots, N-1\}$. We define the subsampling operator $\mathbf{S}: \mathbb{C}^N \mapsto \mathbb{C}^L$ such that, for any $\mathbf{z} \in \mathbb{C}^N$, $(\mathbf{S}\mathbf{z})[j] = \mathbf{z}[i_j]$ for $j \in \{1, \dots, L\}$.
The random Fourier measurements operator is defined, for any $\x \in \R^N$, as
\begin{equation}
    \A\x = \mathbf{S} \mathbf{F} \x := \mathbf{S} \widehat{\x}
    \label{eq:fourier}
\end{equation}
where $\mathbf{F}: \mathbb{C}^N \to \mathbb{C}^N$ is the discrete Fourier transform (DFT) \cite{vetterli2014a}.

We also consider a 2D Laplacian penalty operator $\L_2 = \mathbf{\Delta}$, defined as the 3-points second order discrete derivative (finite differences). We assume periodic boundary conditions, so that the operator $\mathbf{\Delta}$ can be described with a circular convolution \cite{vetterli2014a} with the vector $\mathbf{d}$ as
\begin{equation}
    \forall \x \in \R^N, \qquad \mathbf{\Delta} \x = \mathbf{d} \circledast \x.
    \label{eq:laplacian}
\end{equation}

\begin{lemma}
    The forward matrix $\A = \mathbf{S} \mathbf{F}$ verifies Assumptions~\ref{ass:1}. If the mean of the signal is sampled, meaning $0 \in \mathcal{I}$, then $\A$ and $\L_2 = \mathbf{\Delta}$ also verify Assumptions~\ref{ass:2} and \ref{ass:3}.
\end{lemma}

\begin{proof}
    Due to the DFT properties, the forward matrix satisfies the equality $\A \A^H = N \mathbf{I}_L$. The rows of $\A$ are orthogonal hence linearly independent, so that $\A$ has full row rank and verifies Assumption~\ref{ass:1}.
    It is possible to prove that $\mathrm{ker}\,\mathbf{\Delta} = \mathrm{Span}\left\{ \mathbf{1}_N \right\}$. The nullspace of $\A$ is the vector space spanned by the non-sampled frequencies $\{0, \dots, N-1\} \setminus \mathcal{I}$. If $0 \in \mathcal{I}$, the constant signals do not belong to the nullspace of $\A$ and Assumption~\ref{ass:2} holds. The harmonic DFT basis diagonalizes the convolutions \cite{vetterli2014a}, hence the stability of $\ker{(\A)}^\perp$ is immediately satisfied, verifying Assumption~\ref{ass:3}.
\end{proof}

\begin{proposition}
    If $0 \in \mathcal{I}$, the matrix $\mathbf{M}_{\lambda_2}$ is diagonal and has the following expression 
    \begin{equation*}
        \begin{split}
            \mathbf{M}_{\lambda_2} & = \lambda_2 \diag{\widehat{\mathbf{d}}_\mathcal{I}^2} \left( \mathbf{I}_L + \lambda_2 \diag{ \widehat{ \mathbf{d}}_\mathcal{I}^2 } \right)^{-1} \\
                & = \diag{ \lambda_2 \widehat{ \mathbf{d}}_\mathcal{I}^2 / ( \mathbf{1}_L + \lambda_2 \widehat{ \mathbf{d}}_\mathcal{I}^2 ) }.
        \end{split}
    \end{equation*}
    with $\widehat{\mathbf{d}}_\mathcal{I} = \mathbf{S} \widehat{\mathbf{d}}$.
\end{proposition}

\begin{proof}
    By symmetry of the kernel $\mathbf{d}$, we have $\mathbf{\Delta}^T = \mathbf{\Delta}$.
    Using the circular convolution theorem in Fourier domain \cite{vetterli2014a}, for any vector $\x \in \R^N$ we have $\mathbf{F} \mathbf{\Delta} \x = \mathbf{F}\left(\mathbf{d} \circledast \x\right) = \widehat{\mathbf{d}} \odot \widehat{x} = \diag{\widehat{\mathbf{d}}} \mathbf{F} \x$. Similarly, for $\y \in \mathbb{C}^L$, we can prove $\mathbf{\Delta} \mathbf{F}^H \y = \mathbf{F}^H \diag{\widehat{\mathbf{d}}} \y$.
    Finally
    \begin{equation*}
        \begin{split}
            \A \L_2^T \L_2 \A^H &= \mathbf{S} \mathbf{F} \mathbf{\Delta}^T \mathbf{\Delta} \mathbf{F}^H \mathbf{S}^T \\
                & = \mathbf{S} \diag{\widehat{\mathbf{d}}} \mathbf{F}\mathbf{F}^H \diag{\widehat{\mathbf{d}}} \mathbf{S}^T \\
                & = N \mathbf{S} \diag{\widehat{\mathbf{d}}^2} \mathbf{S}^T \\
                & = N \diag{\widehat{\mathbf{d}}_\mathcal{I}^2}
        \end{split}
    \end{equation*}
    The expression of $\mathbf{M}_{\lambda_2}$ follows by using $\A\A^H = N \mathbf{I}_L$.
\end{proof}

\section{Numerical Simulations}

We simulate a composite inverse problem with forward matrix defined in \eqref{eq:fourier} and solve it using the composite penalty problem  ~\eqref{eq:composite-pbm} with penalty matrices $\L_1 = \mathbf{I}_N$ and $\L_2 = \mathbf{\Delta}$ as defined in \eqref{eq:laplacian}. The reconstructions are performed either in a coupled manner, by directly solving the optimization problem with APGD, or with the decoupled approach suggested by Theorem~\ref{thm:rt}, first solving an $\ell_1$-penalized problem for the sparse component and deducing the smooth component\footnote{Our reproducible code is available at \href{https://github.com/AdriaJ/CompositeSpS}{AdriaJ/CompositeSpS}.}.

\subsection{Simulation model}
\begin{figure}[t]
        \makebox[1.1\columnwidth][c]{
        \begin{subfigure}[b]{\columnwidth}
            \includegraphics[width=1.1\linewidth,right, trim=0 13 0 0, clip]{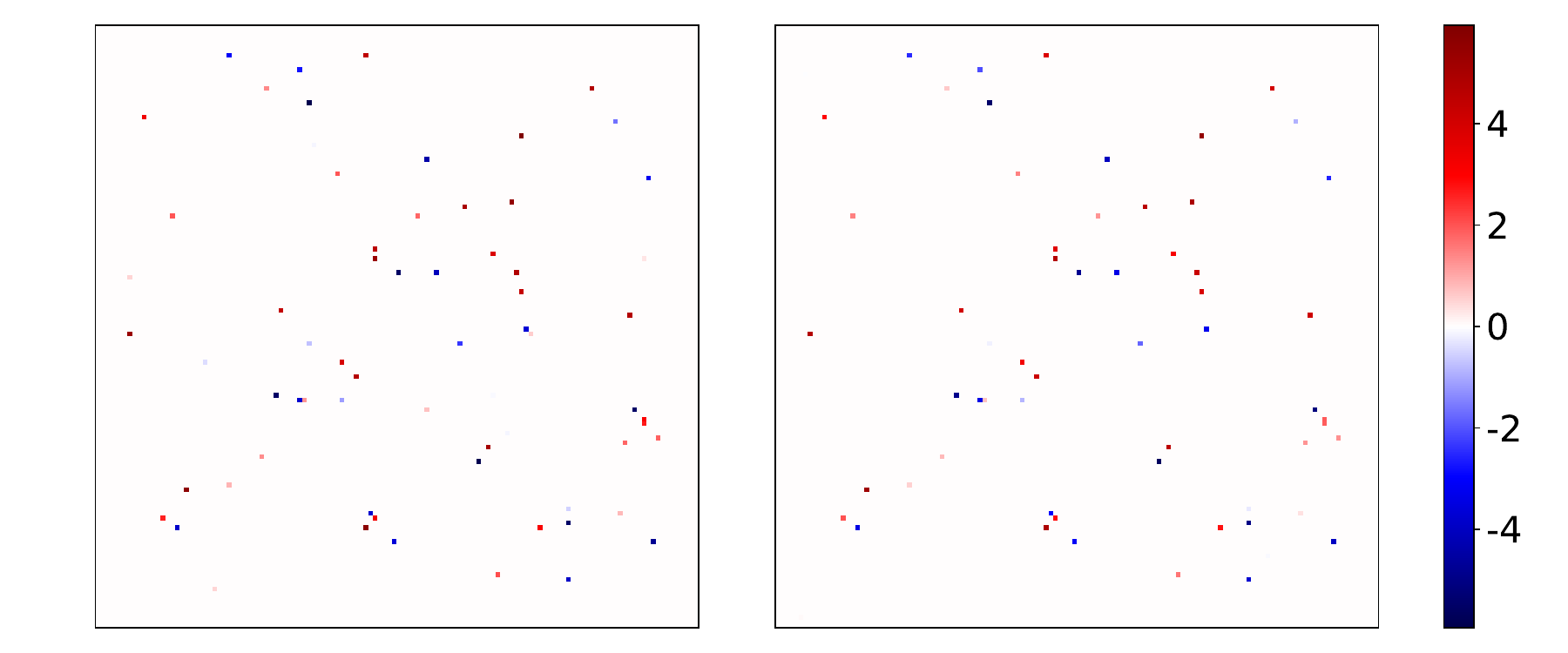}
            \caption{Sparse component $\x_1^*$}
        \end{subfigure}
        }
    \\
        \makebox[1.1\columnwidth][c]{
        \begin{subfigure}[b]{\columnwidth}
            \includegraphics[width=1.1\linewidth,right, trim=0 13 0 -15, clip]{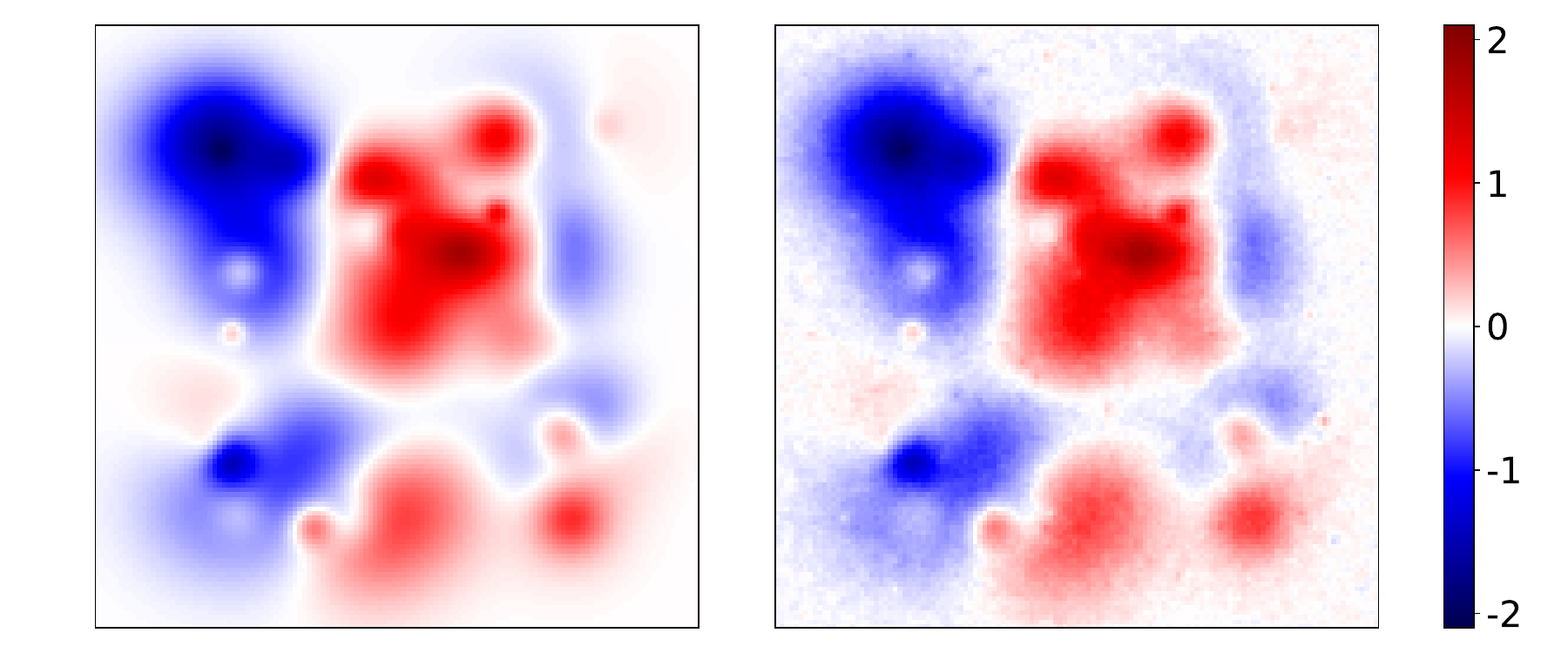}
            \caption{Smooth component $\x_2^*$}
        \end{subfigure}
        }
    \\
        \makebox[1.1\columnwidth][c]{
        \begin{subfigure}[b]{\columnwidth}
            \includegraphics[width=1.1\linewidth,right, trim=0 13 0 -15, clip]{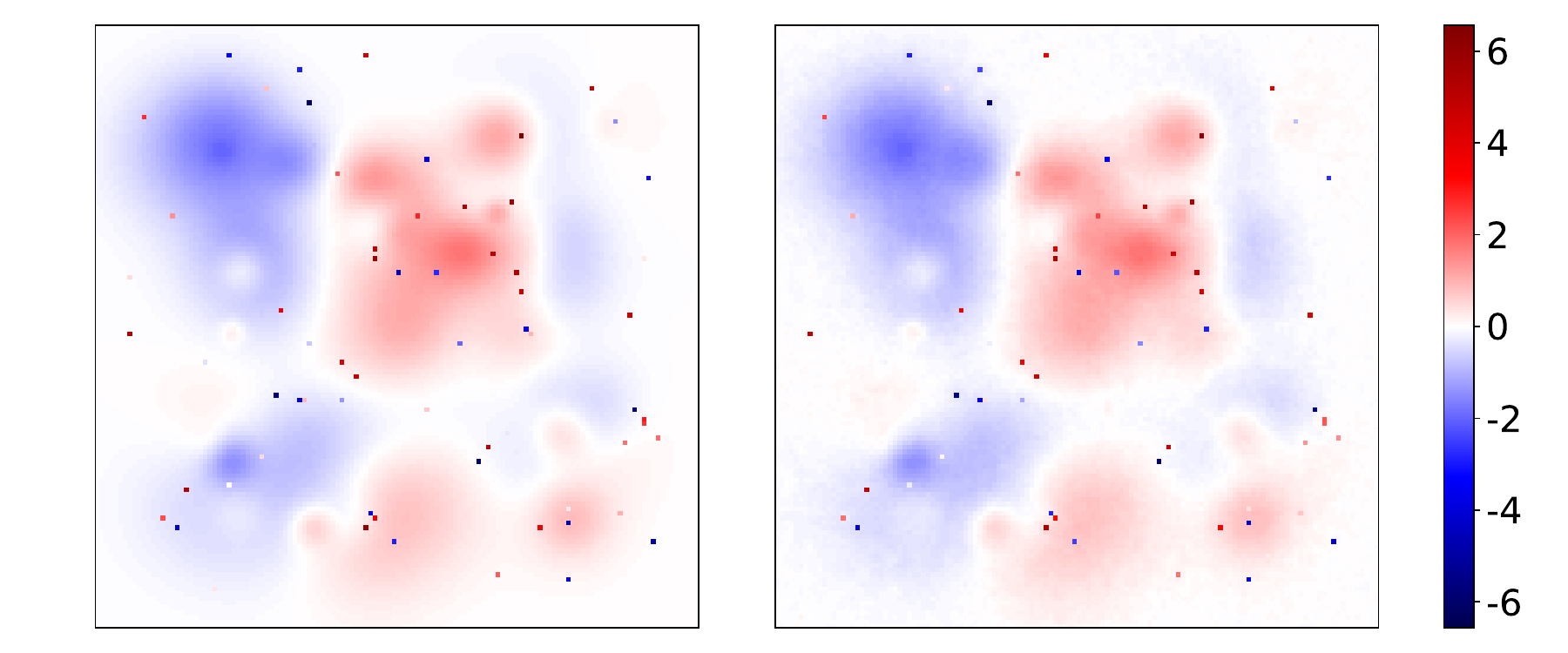}
            \caption{Complete reconstruction $\x_1^* + \x_2^*$}
        \end{subfigure}
        }
    
    \caption{Source (left) and decoupled reconstruction (right) for an image of size $128\times128$ from noisy measurements, $\alpha_1 =0.08$ and $\alpha_2=0.5$.}
    \label{fig:reconstruction}
\end{figure}

The \textit{source image} is simulated as the sum of a sparse component, that is a set of bright isolated pixels, and a smooth component, simulated as a sum of low intensity Gaussian functions with different spatial extensions (left panel of Figure~\ref{fig:reconstruction}). Both the sparse and smooth components take positive and negative values.

The sampled Fourier frequencies $\mathcal{I}$ are drawn at random, half of them following a Gaussian distribution, the other half being uniformly distributed. This heterogeneous model allows to sample both the low and high frequencies of the signal, mimicking the sampling pattern of radio interferometry, which is consistent with the composite signal model. Indeed, the sparse component $\x_1$ presents high frequency components and is thus difficult to reconstruct with low frequency samples only, and conversely for the smooth component $\x_2$. For an image of size $N = n\times m$ pixels, the number of measurements $L$ is chosen as $30$ \% of the number of measurements necessary to exactly reconstruct the image, that is $\sim\!nm/2$. The simulated measurements are finally corrupted with an additive Gaussian white noise with PSNR of $20$dB.

Reconstructions are performed with Python, using the versatile optimization library Pyxu \cite{pyxu-framework}. The coupled method directly solves the problem~\eqref{eq:composite-pbm} of size $2N$ with the APGD solver \cite{liang2017}. The decoupled approach also solves problem~\eqref{eq:decoupled-sparse} of size $N$ with APGD, then the solution of the problem~\eqref{eq:decoupled-smooth} is computed explicitly with its closed-form expression.

\subsection{Parametrization of the regularization parameters}

The regularization parameters are critical  to set and various methods can be used depending on the desired reconstruction characteristics. Theorem~\ref{thm:rt} partly unveils their contribution and relative importance. Therefore, we propose to parameterize $\lambda_1$ and $\lambda_2$ with respectively two parameters $\alpha_1>0$ and $\alpha_2>0$ that still need to be determined according to the noise level and the signal model but no longer depend on the dimensions of the problem.


For the parameter $\lambda_1$, it is known that there exists a maximum value $\lambda_{1, \max}$ above which any solution of a LASSO problem is $\mathbf{0}_N$ \cite{Koulouri_Heins_Burger_2021}. It is common to set the regularization parameter as $\lambda_1 = \alpha_1 \lambda_{1, \max}$
with $0 < \alpha_1 < 1$. Here, the sparse component problem~\eqref{eq:decoupled-sparse} is a simple LASSO problem so that the critical value is given by
\begin{equation*}
    \lambda_{1, \mathrm{max}} = \norm{(\mathbf{M}^{1/2}_{\lambda_2} \A)^H \mathbf{M}^{1/2}_{\lambda_2} \y }_\infty = \norm{\A^H \mathbf{M}_{\lambda_2} \y }_\infty,
\end{equation*}
using the symmetry of the matrix $\mathbf{M}_{\lambda_2}$.

The parameter $\lambda_2$ can be interpreted with an analysis analogous to \cite{hastie2001} (see (3.47), Section 3.4.1). Indeed, $\lambda_2$  modifies the contribution of the reconstruction vectors depending on the associated singular value in the sampling matrix. Here, the smooth component solution $\x^*_2$ involves the matrix $\left(\mathbf{A}^H\mathbf{A} + \lambda_2 \L_2^T \L_2 \right)^{-1}$. We then propose to set $\lambda_2$ so that 
\begin{equation*}
    \lambda_2 \, \sigma_{\max}^2 (\L_2) = \alpha_2 \, \sigma_{\max}^2\left(\A \right),
\end{equation*}
where $\sigma_{\max}$ computes the maximum singular value of a matrix. The parameter $\alpha_2$ controls the intensity of the regularization applied to the forward matrix $\A$, automatically scaling to the actual singular values of $\A$.

Our simulations investigated the different behaviors of decoupled and coupled solving of the composite problem, hence values of $\alpha_1$ and $\alpha_2$ have been handpicked to ensure reasonable reconstructions.

\subsection{Performance analysis}

In terms of images, the solutions produced by the two methods are extremely similar. The decoupled reconstruction is provided in Figure~\ref{fig:reconstruction}. The sparse component was recovered with high accuracy, with a Jaccard index of $0.81$. Despite the grainy texture, the smooth component also recovered the source with good accuracy, displaying a relative $\ell_2$ error of $0.06$ with respect to the source.

\begin{figure}[t]
    \centering
    \includegraphics[width=\linewidth, trim= 0 10 0 20, clip]{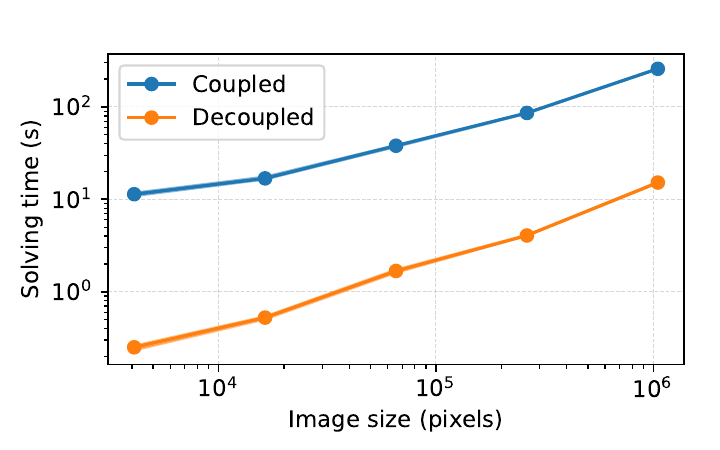}
    \caption{Median reconstruction time over 20 reconstructions for different image sizes. The interquartile spread (shaded) almost coincides with the median.}
    \label{fig:time-bench}
\end{figure}

The most significant benefit of our RT lies in the speed efficiency of the decoupled approach. The solving times of the two methods are reported in Figure~\ref{fig:time-bench}. The decoupled approach was significantly faster, achieving an \textbf{acceleration factor of 17} for the largest images considered and even better in smaller dimensions. This is explained by faster and less numerous iterations. Although the setup considered here may be particularly favorable for a fast decoupled approach, as the matrix $\mathbf{M}_{\lambda_2}$ is diagonal, decoupling the solving of the composite problem \eqref{eq:composite-pbm} has systematically improved the reconstruction times in other simulation setups investigated.


When less measurements $L$ were considered, the reconstruction of the smooth component was less accurate, however we observed some robustness in the estimation of the sparse component. Similarly, when the image size $N$ increases, the grainy texture of $\x_2$ became more visible, irrespective of the the regularization parameter $\alpha_2$. For larger images, it may be beneficial to use a Laplacian kernel with larger support to promote greater smoothness.


\section*{Acknowledgment}

The authors would like to thank Lucas d'Alimonte and Martin Vetterli for help and support. This research was supported by the SNSF with the grant \textit{SESAM - Sensing and Sampling: Theory and Algorithms} (n° 200021\_181978/1.)

\printbibliography

\end{document}